\documentclass[aip,pop,reprint]{revtex4-2}

\usepackage{newtxtext, newtxmath} % font

\usepackage{graphicx} % graphics
\usepackage{hyperref} % hyperlink
\usepackage{siunitx} % to use \micro
\usepackage{mathtools} % \eqref
\usepackage{sidecap} % side caption

\newcommand{\nifs}{National Institute for Fusion Science, 322-6 Oroshicho, Toki, Gifu 509-5292, Japan}
\newcommand{\sokendai}{The Graduate University of Advanced Studies (SOKENDAI), 322-6 Oroshicho, Toki, Gifu 509-5292, Japan}

\begin{document}

\title{Bayesian inference of non-Maxwellian distribution functions from collective Thomson scattering spectra}

\author{Kentaro Sakai}
\email[]{sakai.kentaro@nifs.ac.jp}
\affiliation{\nifs}
\affiliation{\sokendai}

\author{ByungJun Kang}
\affiliation{\nifs}

\author{Akito Nakano}
\affiliation{\nifs}
\affiliation{\sokendai}

\author{Takeo Hoshi}
\affiliation{\nifs}
\affiliation{\sokendai}

\date{\today}

\begin{abstract}

We investigate Bayesian inference of non-Maxwellian electron distribution functions from collective Thomson scattering (CTS) spectra. We represent distribution functions using multiple candidate models of different complexity and perform inference on synthetic spectra generated from known non-Maxwellian distribution functions. Our analysis demonstrates that a sufficiently flexible model can approximate the overall shape of the ground-truth distribution function.
The most plausible model is identified based on model evidence, a statistical measure representing the probability to obtain the observed data given the model. When the candidates include the ground-truth model, the model evidence favors the ground-truth model. Without the ground-truth model as a candidate, the model evidence favors the candidate model with the fewest parameters that adequately approximates the data.
The posterior probability density function of the most plausible model reveals the characteristic features and associated uncertainties of the underlying distribution function. The origin of the additional spectral peaks is attributed to a combination of modifications to the dispersion relation and reduced Landau damping. This enables the objective and data-driven identification of distribution functions directly from observed CTS spectra. 

\end{abstract}

\maketitle

\section{Introduction}
\label{sec_intro}
 
Space and astrophysical plasmas are in non-equilibrium states, where electron and ion velocity distribution functions frequently deviate from a Maxwellian distribution function as the collisional relaxation is negligible. The free energy associated with non-Maxwellian distribution functions dissipates through interactions with collective electromagnetic fields, and as a result, the phenomena can differ significantly from magnetohydrodynamic descriptions \cite{scholer03jgr,matsukiyo06jgr,fujimoto14grl}. To understand the underlying physics, we have been conducting laboratory experiments simulating space and astrophysical plasmas using high-power lasers \cite{takabe21hpl}. Recent advances in experimental and diagnostic techniques enable us to detect signals that suggest the presence of non-Maxwellian distribution functions, including non-thermal tails \cite{fiuza20nphys,yao21nphys} and deformations in bulk distribution functions \cite{morita22pre,zhang23nphys,ji24pop}, and collective electromagnetic fields \cite{sakai22srep}.

Collective Thomson scattering (CTS) is a unique tool to measure local distribution functions of electrons and ions in laser-produced plasmas \cite{froula11}. The conventional CTS analysis usually assumes the plasma to be linear, steady, equilibrium, and stable, i.e., the electron and ion distribution functions are Maxwellian. However, laser-produced plasmas exhibit nonlinear, unsteady, non-equilibrium, and unstable nature, e.g., the thermal conduction by high-energy particles in the presence of temperature gradients generates asymmetric distribution functions \cite{henchen18prl} and the inverse Bremsstrahlung heating selectively heat low-energy particles to form flattop or super-Gaussian distribution functions \cite{turnbull20nphys,milder21prl}. In some cases, the scattered spectrum of non-Maxwellian distribution functions is completely different from that of Maxwellian distribution functions \cite{sato26pop,sakai20pop,sakai23pop}. While the dynamic structure factor, the standard theory of CTS spectra given the electron and ion distribution functions, generally cannot describe CTS spectra in non-equilibrium plasmas, the dynamic structure factor effectively explains CTS spectra in quasi-steady states, such as those after the saturation of instabilities \cite{chapman11prl,sakai20pop,sakai23pop}. 

Several CTS spectra exhibiting peak asymmetry \cite{morita22pre} and additional peaks \cite{zhang23nphys,ji24pop} cannot be explained by a single Maxwellian fit, indicating the presence of non-Maxwellian electron and ion distribution functions. However, identifying the most plausible distribution function remains challenging; the existence of numerous candidate models and the increased degrees of freedom in their parameters often lead to non-unique solutions or failed convergence. In previous studies, the estimated distribution functions are compared with theoretical models or numerical simulations to validate the inferred distribution functions \cite{morita22pre,zhang23nphys,ji24pop}. To characterize distribution functions relying solely on experimental data, a robust data-driven method capable of identifying a unique global optimum and selecting the most plausible model is required.

In this paper, we investigate Bayesian inference of non-Maxwellian distribution functions from CTS spectra. While previous studies primarily employed Bayesian inference as an optimization tool to locate global solutions \cite{schaeffer19prl,foo23aipa,escalona23srep}, we leverage it for both uncertainty quantification and objective model selection. We represent distribution functions as a superposition of several Maxwellian components and evaluate the posterior probability density functions of the underlying parameters. To validate our approach, we construct two ground-truth distribution functions: using a triple-Maxwellian model to replicate the features reported by \citet{zhang23nphys} and a super-Gaussian distribution function observed during laser heating \cite{turnbull20nphys,milder21prl}. The most plausible model is identified based on model evidence, a statistical measure showing the probability to obtain the observed data given the model. The model selection is limited to the candidate models considered and therefore depends on whether an appropriate functional form is included. When the ground-truth model is included among the candidates, the model evidence favors the ground truth. Otherwise, a sufficiently flexible candidate model can provide a reasonable approximation of the overall shape of the ground-truth distribution function, while the model evidence favors the simplest candidate that adequately explains the data.
The posterior probability density functions show that the non-ideal peaks in the triple-Maxwellian spectrum correspond to the non-Maxwellian component of the distribution function. Furthermore, the estimated distribution function and dielectric response function including their associated uncertainties reveal that the emergence of additional peaks is attributed to a combination of modifications to the dispersion relation and reduced Landau damping.

\section{Method}\label{sec:method}

\subsection{Non-Maxwellian model of collective Thomson scattering} \label{sec:model}

To demonstrate the capability of Bayesian inference for non-Maxwellian distribution functions, we perform fittings of scattered spectra assuming various non-Maxwellian models. 
We adopt a multi-Maxwellian model for electrons \cite{sakai20pop,sakai23pop}, where the total electron distribution function $f_e$ is written as a superposition of multiple Maxwellian components; $f_e(v) = \sum_j (n_{e,j} / n_e) f_{e,j} (v)$, where $f_{e,j}$ and $n_{e,j}$ represent the $j$-th component of the Maxwellian distribution function and electron density, respectively. The total electron density satisfies $n_e = \sum_j n_{e,j}$. Each individual component is defined as $f_{e,j}(v) = \exp [-(v-v_{e,j})^{2}/v_{te,j}^2]/(\pi^{0.5} v_{te,j})$, where $v_{e,j}$ and $v_{te,j}$ denote the mean and thermal velocities of the $j$-th component, respectively. This model uses $3j$ parameters for electrons: the density, temperature, and mean velocity for each component.
In this study, we assume a Maxwellian ion velocity distribution function; $f_i(v) = \exp [-(v-v_{i})^{2}/v_{ti}^2]/(\pi^{0.5} v_{ti})$, where $v_i$ and $v_{ti}$ are the mean and thermal velocities of the ions. The thermal velocity is $v_{te,i} = (2 k_b T_{e,i}/m_{e,i})^{0.5}$, where $k_b$ is the Boltzmann constant, $T_{e,i}$ is the temperature, and $m_{e,i}$ is the mass. The dynamic structure factor, which determines the scattered spectra in near-equilibrium plasmas, is given by \cite{froula11}
\begin{equation}
    S(k,\omega) = \frac{2\pi}{k} \left[ \left|1-\frac{\chi_{e}}{\varepsilon}\right|^{2} f_{e} \left(\frac{\omega }{k}\right) + Z \left|\frac{\chi_{e}}{\varepsilon}\right|^{2} f_{i} \left(\frac{\omega }{k}\right) \right],
    \label{eq:form}
\end{equation}
where $k$, $\omega$, and $\chi_{e}$ are the scattering wavenumber, scattering frequency, and electron susceptibility, respectively. The dielectric response function satisfies $\varepsilon = 1 + \chi_e + \chi_i$, where $\chi_i$ is the ion susceptibility. We assume quasi-neutrality such that the ion charge state $Z$ satisfies $q_i n_i + q_e n_e = e (Z n_i - n_e) = 0$, where $n_i$ is the ion density, $q_{e,i}$ is the charge of an electron or ion, and $e$ is the elementary charge. 
The susceptibility for each species is expressed as 
\begin{equation}
\begin{split}
    \chi_{e,i} &= \frac{4\pi q_{e,i}^2 n_{e,i}}{m_{e,i} k^2} \int_{-\infty}^{\infty} \frac{\frac{\partial f_{e,i} (v)}{\partial v}}{\frac{\omega}{k}-v}dv \\
    &= -\frac{4\pi q_{e,i}^2}{m_{e,i}k^2}\sum_{j}\left[\frac{n_{e,i}^{j}}{v_{te,i}^2}Z' \left(\frac{\frac{\omega}{k}-v_{dj}}{v_{te,i}}\right)\right], 
    \label{eq:chi}
\end{split}
\end{equation}
where $Z'(\xi)$ is the derivative of the plasma dispersion function. 
In this study, we focus on the electron feature. For a simple interpretation, we neglect the second term of the right-hand side in Eq.~\eqref{eq:form} and assume $\chi_i \sim 0$. This simplification leads to $S(k,\omega) \sim (2\pi/k) |1/\varepsilon|^{2} f_e(\omega/k)$. The spectral response is determined by $f_e$ and $\varepsilon$. All numerical calculations use the full dynamic structure factor, including the ion susceptibility and ion feature. The wavelength range in which the ion feature appears is removed by the notch filter and excluded from the evaluation of the objective function to be minimized. Therefore, the ion contribution has a negligible effect on the inference even for the lower-velocity electron components.

\begin{figure}
    \includegraphics[width=\hsize]{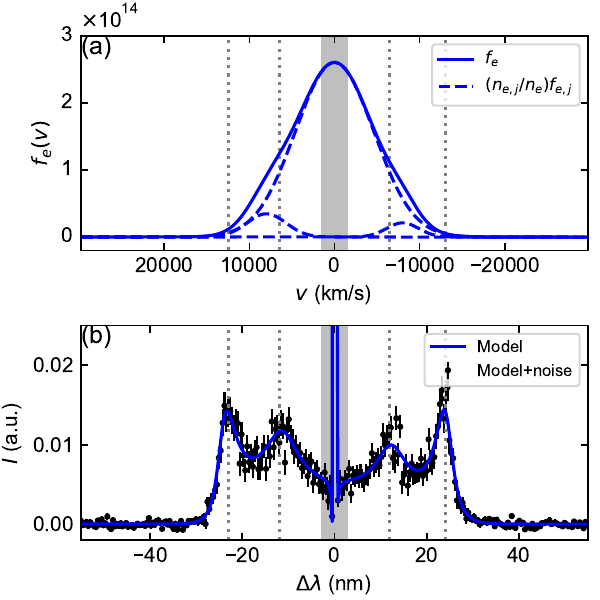}
    \caption{(a) Electron distribution function represented by three Maxwellian components to generate a synthetic spectrum. (b) Synthetic spectrum with noise.}
    \label{fig:answer}
\end{figure}

We generate a synthetic CTS spectrum using a non-Maxwellian electron distribution function consisting of the sum of three Maxwellian distribution functions, replicating the non-Maxwellian features reported in Ref.~\onlinecite{zhang23nphys}. We define the parameters of three Maxwellian components as $(T_{e,j}, n_{e,j}, v_{e,j}) = (120, \SI{3e18}{}, 0)$, $(30, \SI{2e17}{}, 8000)$, and $(\SI{20}{eV}, \SI{1e17}{cm^{-3}}, \SI{-8000}{km/s})$. The solid curve in Fig.~\ref{fig:answer}(a) shows the electron distribution function using the parameters above, while the dashed curves represent the individual Maxwellian components. 
Figure~\ref{fig:answer}(b) shows the synthetic spectrum obtained using the electron distribution function in Fig.~\ref{fig:answer}(a), plotted as a function of the wavelength change $\Delta \lambda = \lambda - \lambda_0$. We set the measurement configuration with a probe laser wavelength of $\lambda_0 = \SI{526.5}{nm}$ and a scattering angle of $\theta=\SI{63}{\degree}$. We generate the synthetic spectrum over the wavelength range $-100\le \Delta\lambda \le \SI{100}{nm}$. To reproduce the realistic experimental conditions, we adopt a wavelength axis with a wavelength interval of \SI{0.41}{nm}, add noise to the model spectrum, and remove the data at $-3 \le \Delta \lambda \le \SI{3}{nm}$ as a notch filter. The solid curve represents the model spectrum without noise. The gray background corresponds to the wavelength range of the notch filter, which is also indicated in Fig.~\ref{fig:answer}(a). We incorporate both Poisson noise arising from photon statistics and Gaussian noise representing both dark current and readout noise. We evaluate the magnitude of the Poisson noise based on the square root of the counts, assuming a total photon count of $10^{4}$. We set the standard deviation of the Gaussian noise in each wavelength channel to half the square root of the mean photon count, incorporating a 20\% random fluctuation across channels. The total uncertainty is calculated as the root sum square of these independent noise components. The spectral intensity is normalized such that the sum of all intensities is unity. Hereinafter, we refer to the black markers as the inference data. Figures~\ref{fig:answer}(a) and \ref{fig:answer}(b) are made equivalent to account for the Doppler shift; $\Delta \lambda \simeq - 2 \lambda_0 (v/c) \sin (\theta/2)$, where $c$ is the speed of light. The spectrum exhibits four distinct peaks at $\Delta \lambda \sim -23, -12, 12, \SI{24}{nm}$. The peak wavelengths and associated phase velocities are indicated by vertical dotted lines in Figs.~\ref{fig:answer}(b) and \ref{fig:answer}(a), respectively. Although the peaks at $\Delta \lambda \sim -23, \SI{24}{nm}$ are considered to be the electron plasma wave resonance, the other peaks at $\Delta \lambda \sim -12, \SI{12}{nm}$ are not explained by the resonant condition of any wave modes assuming a single Maxwellian distribution function. 
The associated phase velocities of the additional peaks are $v\sim 6500, \SI{-6500}{km/s}$. Since the electron distribution function is single-humped and has no local minima, the plasma is expected to be stable against electrostatic instabilities \cite{penrose60pof}. In Appendix~\ref{sec:dispersion}, we calculate the linear dispersion relation using the non-Maxwellian electron distribution function in Fig.~\ref{fig:answer}(a) to discuss the stability and the associated wave modes in more detail.

\subsection{Bayesian inference}\label{sec:bayes}

Bayes’ theorem relates the posterior probability of the parameters given the data [$P(X|Y)$] to the likelihood of the data given the parameters [$P(Y|X)$], the prior probability [$P(X)$], and the model evidence [$P(Y)$] as 
\begin{equation}
    P(X|Y) = \frac{P(Y|X) P(X)}{P(Y)}, 
    \label{eq:bayes}
\end{equation}
where $X$ represents the set of parameters and $Y$ denotes the observed data. 
Monte Carlo sampling provides a general framework for explicitly quantifying uncertainty by representing the Bayesian posterior probability distribution function as a multidimensional histogram constructed from a large number of samples. In conventional Markov chain Monte Carlo methods, samples are generated sequentially along a Markov chain. This inherently sequential nature limits the degree of parallelism and makes it difficult to fully exploit the computational capabilities of modern massively parallel architectures. In contrast, population annealing Monte Carlo (PAMC) evolves a large population of samples simultaneously through an annealing process \cite{hukushima03aipp}. Since the samples can be generated and updated in parallel, PAMC is inherently well suited to massively parallel computation and can efficiently exploit the computational resources of modern supercomputers.
We sample the posterior probability density function using the PAMC method implemented in the data analysis software ODAT-SE (formerly known as 2DMAT) \cite{motoyama22cpc,sakai26ppcf}. For the numerical parameters of the PAMC sampling, we use 3200 replicas and perform 5000 sampling steps before an annealing.
We perform the Monte Carlo sampling within a $3j$-dimensional parameter space. The sampling region for the electron temperature, density, and velocity for each Maxwellian component are $10\le T_{e,j} \le \SI{500}{eV}$, $0\le n_{e,j} \le \SI{5e18}{cm^{-3}}$, and $\SI{-10000}{}\le v_{e,j} \le \SI{10000}{km/s}$, respectively. Within these regions, we adopt a uniform prior probability for all parameters. By definition,
\begin{equation}
    P(X) = \frac{1}{\int dX}
    \label{eq:prior}
\end{equation}
and is a constant.

We define the objective function to be minimized as
\begin{equation}
    F(X;Y) = \sum_n^N \left[ \frac{I_{data}(\lambda_n) - I_{model}(\lambda_n; X)}{\sigma(\lambda_n)} \right]^2, 
    \label{eq:R2}
\end{equation}
where $I_{data}(\lambda_n)$ and $\sigma(\lambda_n)$ represent the observed spectral intensities and their associated uncertainties, respectively. Here, the parameter set is $X = (T_{e,1},n_{e,1},v_{e,1}, \dots, T_{e,j},n_{e,j},v_{e,j})$, and the observed data is $Y = [I_{data}(\lambda_1),\dots, I_{data}(\lambda_N)]$. We calculate the spectral intensity of a model $I_{model}(\lambda_n; X)$ using the dynamic structure factor defined in Eq.~\eqref{eq:form} for wavelength data points $\lambda_n$. 
The multi-Maxwellian model introduced in Sec.~\ref{sec:model} can yield distribution functions unstable to electrostatic instabilities. During the Bayesian inference, we effectively exclude unstable solutions by assigning a sufficiently large value to $F(X;Y)$ when the Penrose criterion is satisfied \cite{penrose60pof}.
In this analysis, we select the wavelength ranges $-80\le \Delta \lambda\le \SI{-3}{nm}$ and $3\le \Delta \lambda\le \SI{80}{nm}$, which includes $N=375$ data points. Around the most plausible parameters, $F(X;Y)/N$ approaches unity because the residual scales with the uncertainty; $I_{data}(\lambda_n) - I_{model}(\lambda_n; X) \sim \sigma(\lambda_n)$, which is one of several indicators to find the most plausible model.

Although photon counting noise is intrinsically described by Poisson statistics, the number of detected photons considered here is sufficiently large for the Poisson distribution to be accurately approximated by a Gaussian distribution. 
The likelihood for an individual data point is given by
\begin{equation}
\begin{split}
    &P(Y_n|X;\beta) = \\
    &\left[ \frac{\beta}{\pi \sigma(\lambda_n)^2}\right]^{\frac{1}{2}}
    \exp \left[-\beta \frac{\{I_{data}(\lambda_n) - I_{model}(\lambda_n; X)\}^2}{\sigma(\lambda_n)^2}\right]. 
    \label{eq:likelihood_single}
\end{split}
\end{equation}
Consequently, the total likelihood for the entire dataset becomes the product of these individual likelihoods in Eq.~\eqref{eq:likelihood_single}; 
\begin{equation}
\begin{split}
    P(Y|X;\beta) &= \prod_n^N P(Y_n|X;\beta) \\
    &= \left(\frac{\beta}{\pi}\right)^{\frac{N}{2}} \left[\prod_n^N \frac{1}{\sigma(\lambda_n)}\right] \exp [-\beta F(X;Y)]. 
    \label{eq:likelihood}
\end{split}
\end{equation}
The hyperparameter $\beta$ represents the degree of uncertainty in the context of the annealing process, effectively controlling the noise level. In the PAMC runs, we start with $\beta =0$ and gradually increase $\beta$ up to 1. Unlike some studies \cite{escalona23srep} that optimize $\beta$ by including it in the parameter set $X$, we treat $\beta$ as a hyperparameter to facilitate efficient exploration of the high-dimensional parameter space. 
To ensure the robustness of our results, we performed PAMC runs for electron distribution models consisting of one to four Maxwellian components. For each model, we performed three independent runs with different random seeds to verify statistical consistency and reproducibility. Using 800~cores of the Fugaku supercomputer, a single PAMC run requires 2--3~hours, although the computation time depends on the model, the number of parameters, and the available computational resources.

\section{Results}\label{sec:result}

\subsection{Comparison among models}\label{sec:comparison}

\begin{figure*}
    \includegraphics[width=\hsize]{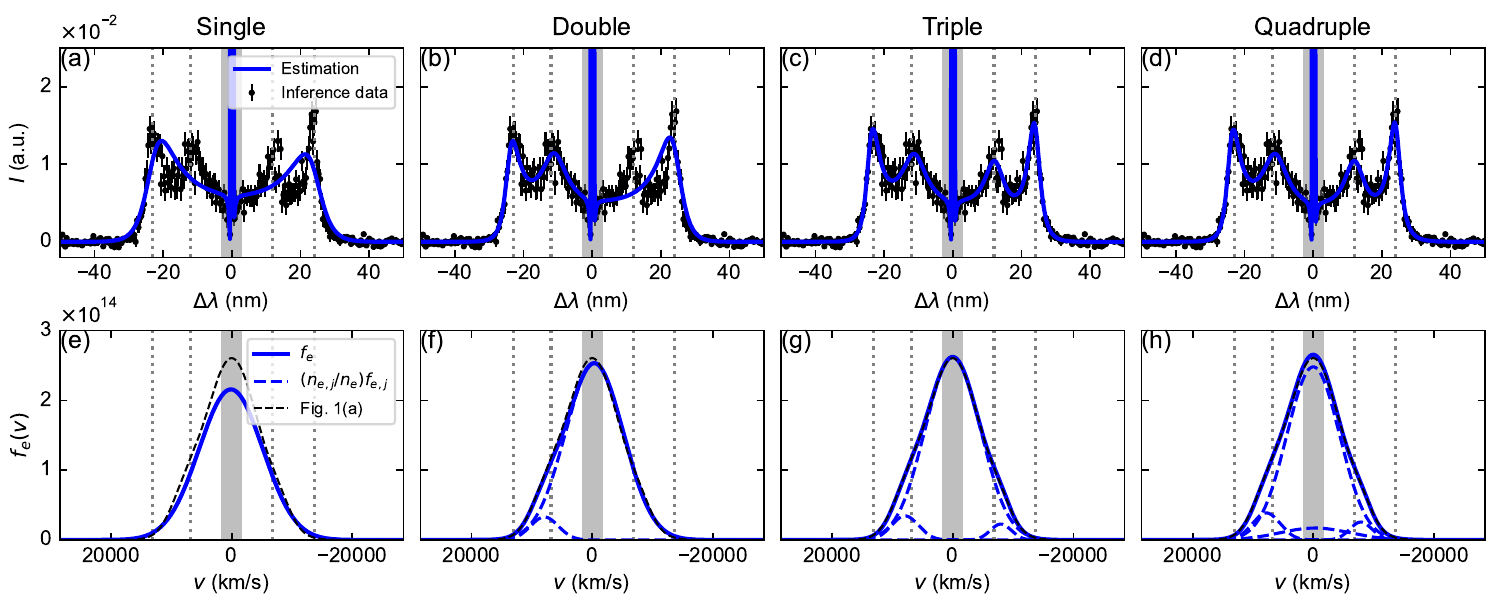}
    \caption{Comparison of the results with the smallest objective function value for the (a,e) single, (b,f) double, (c,g) triple, and (d,h) quadruple Maxwellian models using the spectrum shown in Fig.~\ref{fig:answer}(b). (a--d) Best-fit spectra for the four models. (e--h) Estimated electron distribution functions.}
    \label{fig:comparison}
\end{figure*}

Figure~\ref{fig:comparison} compares the best-fit results with the smallest objective function value among the single, double, triple, and quadruple models. We plot the optimized model spectra as blue curves in Figs.~\ref{fig:comparison}(a)--\ref{fig:comparison}(d), together with the inference data as black markers. The gray background and vertical dotted lines remain consistent with the definitions in Fig.~\ref{fig:answer}(b). 
The inferred distribution function more closely resembles the ground-truth input as the number of Maxwellian components increases. Specifically, the triple and quadruple models successfully capture the two additional spectral peaks at $\Delta \lambda \sim -12$ and \SI{12}{nm}, whereas the single and double models fail to reproduce these features. The minimum objective function values in the single, double, triple, and quadruple models are $\min F(X;Y)/N \sim 2.1$, 1.5, 0.95, and 0.94, respectively. Since $F(X;Y)/N \sim 1$ for the triple and quadruple models, the triple and quadruple models explain the inference data within the noise level. In contrast, the single and double models provide inadequate descriptions of the observations. 
The transition from the triple to the quadruple model yields no significant improvement, suggesting that three components are sufficient to represent the underlying physics.
In the single model, the peaks at $\Delta \lambda \sim -23$ and \SI{24}{nm} are fitted, while those at $\Delta \lambda \sim -12$ and \SI{12}{nm} are not. The spectral shape shows two electron plasma wave resonant peaks, but the remaining two peaks are not explained by the single model. In the double model, the peaks at $\Delta \lambda \sim -23$, $-12$, and \SI{24}{nm} are fitted, while that at $\Delta \lambda \sim \SI{12}{nm}$ is not. The discrepancy highlights the necessity of at least three Maxwellian components to provide a complete, data-driven reconstruction of the non-ideal spectral features observed in this system.

Figures~\ref{fig:comparison}(e)--\ref{fig:comparison}(h) show the electron distribution functions estimated by the fittings in Figs.~\ref{fig:comparison}(a)--\ref{fig:comparison}(d), respectively. The blue-solid and blue-dashed curves represent the total electron distribution function and its Maxwellian component, respectively. The ground-truth electron distribution function in Fig.~\ref{fig:answer}(a) is shown in black-dashed curves. The gray background and vertical dotted lines are the same as those in Fig.~\ref{fig:answer}(a). 
Comparing the blue-solid and black-dashed curves, we find that the distribution functions inferred from the triple and quadruple models closely reproduce the ground-truth electron distribution function. While the distribution function inferred from the double model appears close to the ground-truth electron distribution function, its deviation from the ground truth is larger than those inferred from the triple and quadruple models.
In the triple and quadruple models, two components have mean velocities comparable to the phase velocities of additional spectral peaks, while the remaining component(s) exhibit the mean velocity close to zero. The sum of two near-zero-velocity components in the quadruple model effectively matches the single near-zero component in the triple model, indicating that the triple model is effectively equivalent to the quadruple model. Therefore, based on the qualitative examination of the inferred velocity distribution functions, the triple model appears to be the most plausible.

\subsection{Model selection}\label{sec:model_evidence}

The comparison of four analyses in Sec.~\ref{sec:comparison} suggests that the triple model is considered to be the most plausible, primarily because we can directly compare the estimated electron distribution function with the known ground truth in Fig.~\ref{fig:answer}(a). However, in the analysis of experimental data, the ``true'' electron distribution function is generally unknown unless independent diagnostics provide verification. This casts doubt on the validity of multi-Maxwellian analyses applied to experimental data unless the estimated distribution function is supported by theoretical models or numerical simulations \cite{morita22pre,zhang23nphys}. 

We identify the most plausible model among the four candidates considered here through an objective and data-driven approach that eliminates arbitrariness. To determine the most plausible model, we calculate a statistical measure of model evidence, which is the normalization factor of the posterior probability density function and represents the probability of obtaining the data under a given model \cite{friel12sn}. 
The model evidence is calculated by integrating Eq.~\eqref{eq:bayes} over the entire parameter space. Using Eqs.~\eqref{eq:prior} and \eqref{eq:likelihood}, we obtain 
\begin{equation}
\begin{split}
    P(Y;\beta) &= \frac{\int P(Y|X;\beta) P(X) dX}{\int P(X|Y) dX} \\
    &= \left(\frac{\beta}{\pi}\right)^{\frac{N}{2}} \left[\prod_n^N \frac{1}{\sigma(\lambda_n)}\right] \frac{\int \exp[-\beta F(X;Y)] dX}{\int dX}. 
    \label{eq:model_evidence}
\end{split}
\end{equation}
Note that $\int P(X|Y) dX = 1$ by the definition of a probability density function. The ratio of integrals on the right-hand side is calculated within the PAMC method implemented in ODAT-SE with the initial hyperparameter set to $\beta=0$ \cite{motoyama22cpc}. 
Assuming that the integral $\int \exp[-\beta F(X;Y)] dX$ is dominated by the neighbourhood of the minimizer $X^\ast$ of $F(X;Y)$, the ratio of integrals is approximately proportional to $\exp[-\beta F(X^\ast;Y)]$, and the condition of $dP/(d\beta)=0$ gives the optimized $\beta^\ast \sim N/[2 F(X^\ast;Y)]$. For a proper fit, we expect $F(X^\ast;Y)/N\sim 1$ and an optimized $\beta^\ast\sim 0.5$. This condition holds when the residuals, $I_{data}(\lambda_n)-I_{model}(\lambda_n;X^\ast)$, follow a Gaussian distribution with a standard deviation of $\sigma(\lambda_n)$ and the number of data points is sufficiently large ($N\gg1$). The condition $\beta^\ast\sim0.5$ therefore provides another indicator, in addition to $\min F(X;Y)/N\sim1$, that the model reproduces the data within the assumed noise level. In this study, we generate the inference data by adding Gaussian noise with the standard deviation proportional to the signal intensity. Given that the number of data points much greater than unity, the expectations $F(X^\ast;Y)/N\sim 1$ and $\beta^\ast\sim 0.5$ must be satisfied. In the case of underfitting, the objective function exceeds the noise level [$F(X^\ast;Y)/N>1$], resulting in $\beta^\ast<0.5$. Conversely, overfitting yields the reverse relationship: $F(X^\ast;Y)/N<1$ and $\beta^\ast>0.5$.

\begin{figure}
    \includegraphics[width=\hsize]{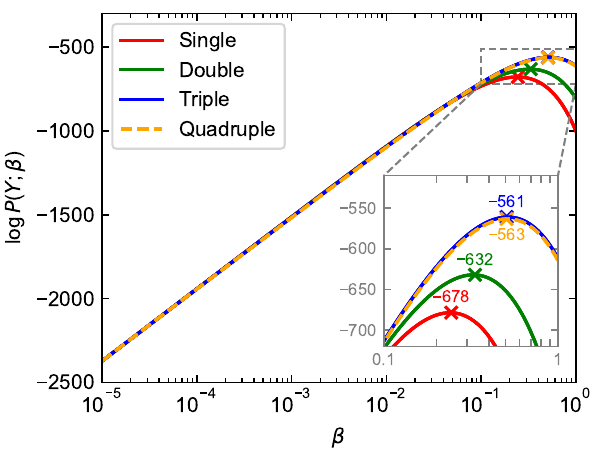}
    \caption{Model evidence as a function of $\beta$ for the four models considered in the analysis of the spectrum shown in Fig.~\ref{fig:answer}(b). Twelve curves represent three independent runs with different random seeds for each of the four models; both statistical uncertainty and random-seed dependence are negligible on the scale of the figure.}
    \label{fig:model_evidence}
\end{figure}
Figure~\ref{fig:model_evidence} shows the model evidence for each of the four models as a function of the hyperparameter $\beta$. To ensure statistical robustness, we plot the results from the three independent PAMC runs with different random seeds in the same color for each model. The model evidence curves for each model obtained with different random seeds are indistinguishable on the scale of the figure. The variance across different seeds is approximately 0.01, which is much smaller than the differences between the models themselves, indicating that the statistical uncertainty among independent runs is negligible and that the model selection result presented below is statistically robust. 
At low $\beta$, the model-evidence curves exhibit power-law behavior and are nearly identical across all models. This behavior is explained by Eq.~\eqref{eq:model_evidence}; the exponential term approaches unity at low $\beta$ ($\exp [\beta F(X;Y)] \sim 1$), causing the likelihood functions for all models to converge to the identical value of $(\beta/\pi)^{N/2} [\prod_n^N \sigma(\lambda_n)^{-1}]$. However, the curves begin to diverge at $\beta \gtrsim 0.1$. The inset in Fig.~\ref{fig:model_evidence} shows a magnified view of the region $0.1\le\beta\le 1$ and $-720\le \log P(Y;\beta) \le -510$, indicated by the dashed rectangle. The peaks of the model evidence curves are marked along with their peak values. Since the peak of the model evidence corresponds to the optimal $\beta^\ast$ for a given model, comparison of peak values allows us to objectively identify the most plausible model among the four candidates considered here. The model evidence for the single and double models peaks at lower $\beta$ values compared to the triple and quadruple models. The $\beta^\ast$ values are lower than 0.5, indicating that the lower-order models are underfitting the data. Since $\beta^{-1}$ represents the noise level, discrepancies between the estimated spectrum and the inference data that cannot be explained by the model are reflected in the value of $\beta$ in these analyses. Because the model evidence peaks at $\beta^\ast\sim 0.5$ in the triple and quadruple models, both models fit the inference data well. Therefore, the Monte Carlo sampling achieves sufficient statistical accuracy in the triple and quadruple models, yielding results that are consistent with the generation process of the inference data. 
The peak value for the triple model [$P_3(Y;\beta^\ast)$] is significantly larger than that for the quadruple model [$P_4(Y;\beta^\ast)$], as demonstrated by the relation $\log P_3(Y;\beta^\ast) \sim \log P_4(Y;\beta^\ast) +2$. This indicates that the triple model explains the inference data more effectively. The difference in peak values originates from the $\int dX$ term in Eq.~\eqref{eq:model_evidence}, which increases as the parameter space expands, i.e., with a larger number of Maxwellian components. Consequently, this term penalizes models with higher degrees of freedom, effectively preventing overfitting. Furthermore, the difference is independent of $\beta$; the nearly constant shift between the model evidence curves for the triple and quadruple models in the inset of Fig.~\ref{fig:model_evidence} is consistent with the likelihood function in Eq.~\eqref{eq:likelihood} and the resultant model evidence in Eq.~\eqref{eq:model_evidence}. Because the model-evidence-based evaluation involves no subjective parameters, this approach enables the identification of the most plausible model in a fully data-driven manner.

\subsection{Posterior probability density function}\label{sec:posterior}

\begin{figure*}
    \includegraphics[width=\hsize]{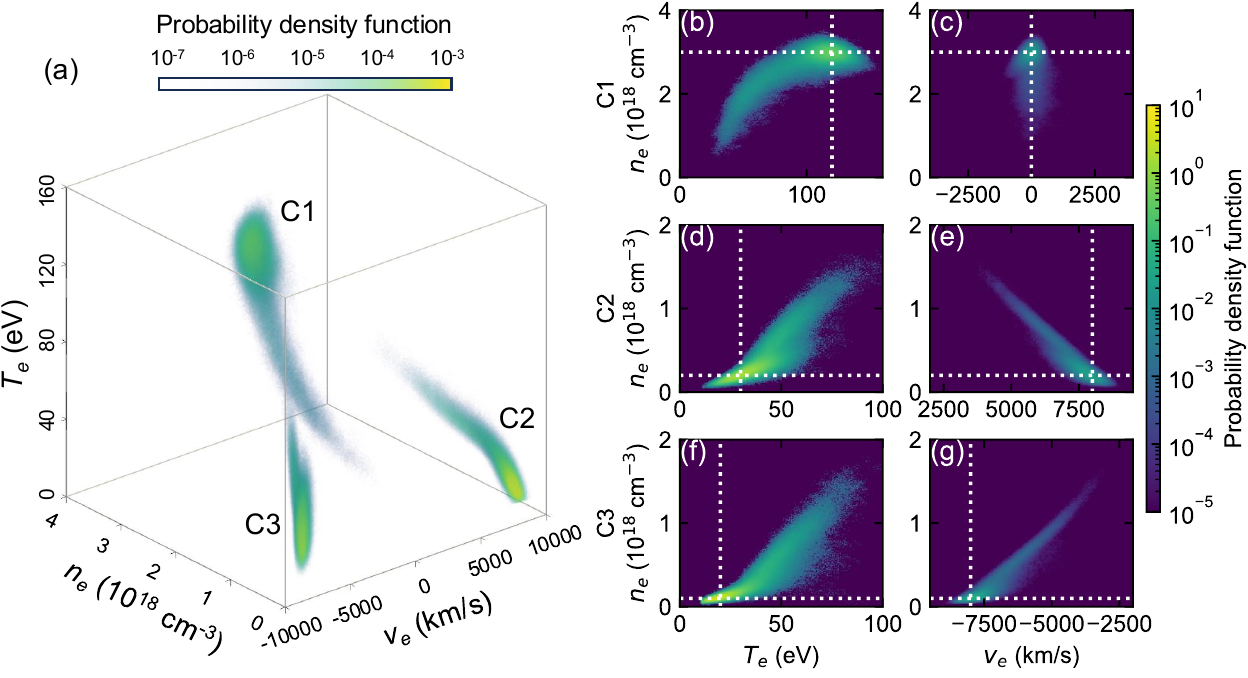}
    \caption{(a) Three-dimensional posterior probability density function of the triple model at $\beta = 0.1$ in the analysis of the spectrum shown in Fig.~\ref{fig:answer}(b). Two-dimensional projected posterior probability density functions of (b,c) C1, (d,e) C2, and (f,g) C3 components.}
    \label{fig:pdf}
\end{figure*}

One of the advantages of Bayesian inference is the ability to visualize parameter uncertainty through the posterior probability density function. Because different combinations of components can produce similar CTS spectra, the inverse problem can generally have multiple solutions. Bayesian inference provides the full posterior probability density function rather than only a single best-fit solution. Multiple distinct solutions consistent with the observed spectrum appear as separate peaks in the posterior probability density function. These solutions can be assessed using physical constraints and prior knowledge to identify the most physically plausible solution.
We plot the three-dimensional posterior probability density function of estimated parameters at $\beta=0.1$ in Fig.~\ref{fig:pdf}(a). Although the selected $\beta$ is below the most probable value ($\beta \sim 0.5$), it provides useful insights into the structures of the parameter space.  
One can see three isolated structures corresponding to the posterior probability density function of each Maxwellian component. The most probable parameters of these isolated structures correspond to the input parameters of three Maxwellian components. The isolated structures show elongated shapes, indicating the high-dimensional correlation of parameters. The parameter space exhibits no spurious local maxima (or local minima in the objective function) given the inference data in Fig.~\ref{fig:answer}. For clarity, we label the component near $v_e\sim 0$ as C1, and the components with positive and negative $v_e$ as C2 and C3, respectively. 

Figures~\ref{fig:pdf}(b)--\ref{fig:pdf}(g) present two-dimensional projections of the three-dimensional posterior probability density function for the C1--C3 components labeled in Fig.~\ref{fig:pdf}(a). These components are extracted by sorting the velocities of the three components and relabeling them, from largest to smallest, as C2, C1, and C3. The white dotted lines indicate the parameters of three Maxwellian components used to generate the inference data. Each peak of the posterior probability density function aligns closely with the white dotted lines, indicating that the ``true'' parameters are estimated. Notably, the posterior probability density function for C1 exhibits a tail elongated toward the low-density and low-temperature region, whereas the tails for C2 and C3 extend toward the high-density, high-temperature, and low-speed regions. Given that the spectral intensity is proportional to the distribution function and its velocity derivative \cite{sakai20pop}, these elongated structures seem to reflect the conservation of both quantities at the corresponding velocities.

\subsection{Inference beyond the candidate model class}\label{sec:superGaussian}

\begin{figure}
    \includegraphics[width=\hsize]{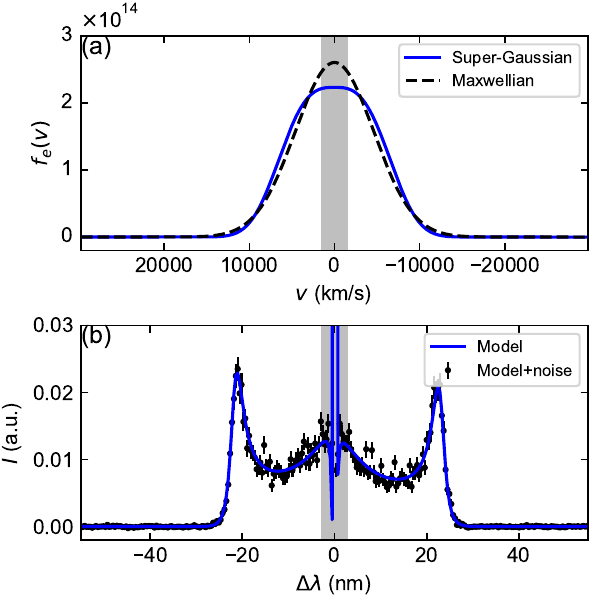}
    \caption{(a) Electron distribution function represented by a super-Gaussian function to generate a synthetic spectrum. (b) Synthetic spectrum with noise.}
    \label{fig:answer_sg}
\end{figure}

In the analyses presented above, we used a synthetic spectrum generated with one of the candidate models. However, in the analysis of experimental data, we generally do not know the exact functional form of the distribution function. To examine how the multi-Maxwellian models perform when none of them exactly represents the underlying distribution function, we generated a synthetic spectrum from a super-Gaussian distribution function \cite{turnbull20nphys,milder21prl} and analyzed the spectrum using the single, double, triple, and quadruple Maxwellian models. We also included the super-Gaussian model as a reference.
As shown in Fig.~\ref{fig:answer_sg}(a), we assume a super-Gaussian distribution function with $m=3$, $T_e=\SI{120}{eV}$, $n_e=\SI{3e18}{cm^{-3}}$, and $v_e=0$. The dashed curve represents a Maxwellian electron distribution function with the same $T_e$, $n_e$, and $v_e$. 
Figure~\ref{fig:answer_sg}(b) shows the synthetic spectrum obtained using the super-Gaussian electron distribution function in Fig.~\ref{fig:answer_sg}(a). We used the same measurement configuration and parameters as in Fig.~\ref{fig:answer}(b). In addition to the two electron plasma wave resonant peaks at $\Delta \lambda \sim -21, \SI{22}{nm}$, the spectrum exhibits a broad local maximum around $\Delta\lambda\sim 0$. The broad central feature is characteristic of CTS spectra from super-Gaussian electron distribution functions \cite{turnbull20nphys,milder21prl}.

\begin{figure*}
    \includegraphics[width=\hsize]{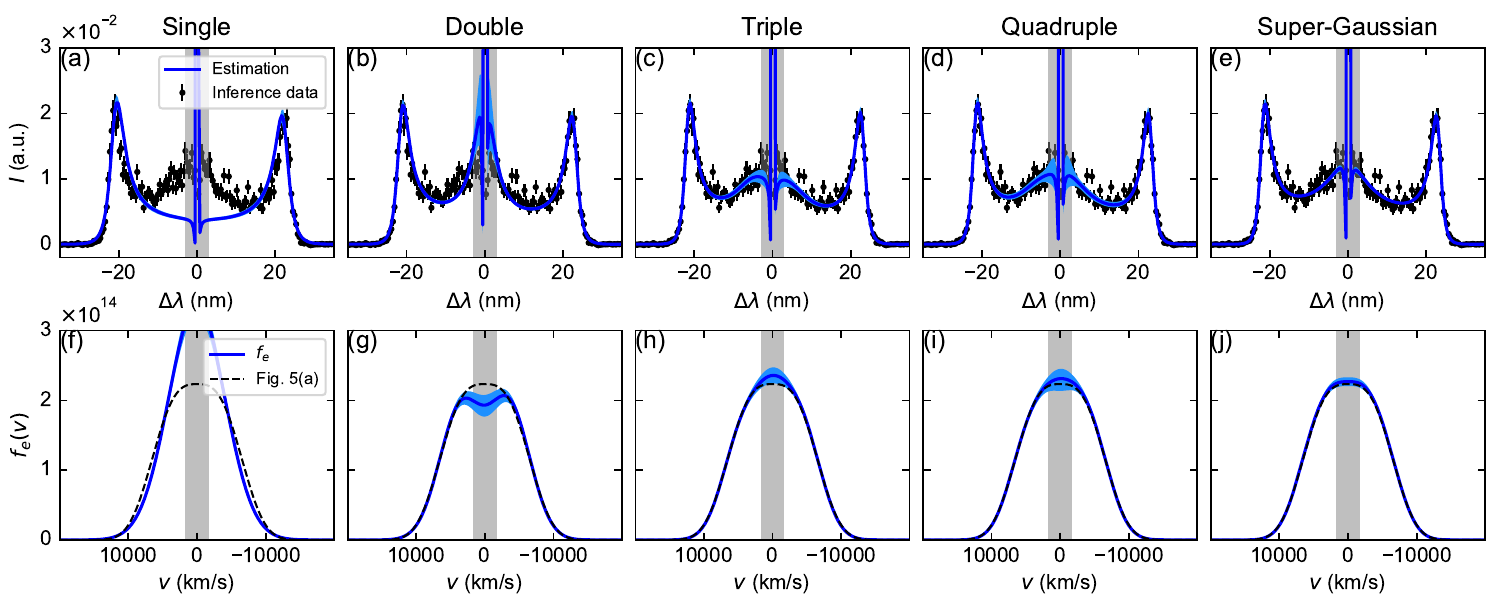}
    \caption{Comparison of the results for the (a,f) single, (b,g) double, (c,h) triple, (d,i) quadruple Maxwellian models and the (e,j) super-Gaussian model using the spectrum shown in Fig.~\ref{fig:answer_sg}(b). (a--e) Fitted spectra. (f--j) Estimated electron distribution functions.}
    \label{fig:comp_fit_sg}
\end{figure*}

Figure~\ref{fig:comp_fit_sg} compares the fitting results for the single, double, triple, and quadruple Maxwellian models and the super-Gaussian model. We plot the 95\% confidence interval for the fitted spectra in Figs.~\ref{fig:comp_fit_sg}(a)--\ref{fig:comp_fit_sg}(e), together with the inference data as black markers. These intervals are determined by extracting the 2.5th and 97.5th percentiles from the ensemble of spectral intensities generated by all replicas at each wavelength data point. For each model, we obtain the confidence interval at the value of $\beta$ that maximizes the model evidence. The gray background remains consistent with the definition in Fig.~\ref{fig:answer_sg}(b). 
The single model reproduces only the two electron plasma wave resonant peaks, whereas the models with two or more Maxwellian components also reproduce the broad local maximum around $\Delta \lambda\sim 0$. For the triple and quadruple models, the fitted spectra show no systematic deviations from the inference data. This improvement with the number of components reflects the flexibility of the multi-Maxwellian model, which can approximate arbitrary velocity distribution functions in the limit of infinitely many components \cite{pavan11pop}. For the more flexible multi-Maxwellian models, the uncertainty in the fitted spectra increases around $\Delta\lambda\sim0$ because this wavelength range is excluded from the inference due to the notch filter. In contrast, the super-Gaussian model shows less uncertainty in this wavelength range because the specific functional form imposes stronger constraints on the spectral shape even in the absence of data. The minimum objective function values are $\min F(X;Y)/N \sim 4.1$, 1.4, 1.07, 1.05 and 1.05 for the single, double, triple, and quadruple Maxwellian models and the super-Gaussian model, respectively.

Figures~\ref{fig:comp_fit_sg}(f)--\ref{fig:comp_fit_sg}(j) show the 95\% confidence interval for electron distribution functions corresponding to the fitted spectra in Figs.~\ref{fig:comp_fit_sg}(a)--\ref{fig:comp_fit_sg}(e), respectively. The black dashed curves indicate the ground-truth distribution function shown in Fig.~\ref{fig:answer_sg}(a). The gray background is the same as that in Fig.~\ref{fig:answer_sg}(a). 
The distribution functions inferred from the single and double models deviate from the ground truth, mainly around $v\sim0$, although the deviation is smaller for the double model. This velocity range corresponds to the broad local maximum around $\Delta\lambda\sim0$, a characteristic feature of CTS spectra from super-Gaussian electron distribution functions. At $|v|\gtrsim \SI{3000}{km/s}$, the models with two or more Maxwellian components capture the overall shape of the ground truth. The distribution functions inferred from the triple and quadruple models agree with the ground truth within their uncertainties. The super-Gaussian model yields the closest agreement with the ground truth and the smallest uncertainty, suggesting that this model is the most plausible of the five candidates.

\begin{figure}
    \includegraphics[width=\hsize]{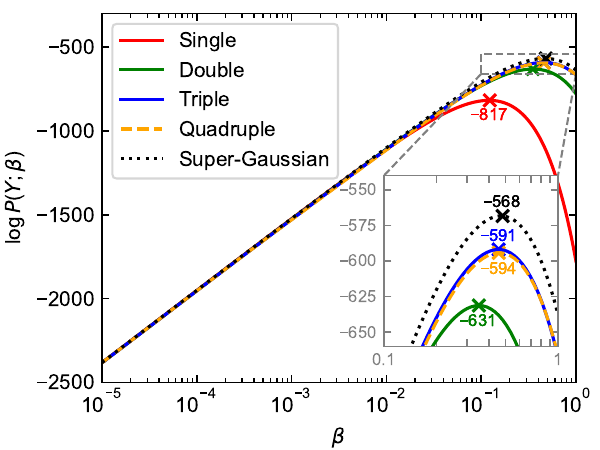}
    \caption{Model evidence as a function of $\beta$ for the five models considered in the analysis of the spectrum shown in Fig.~\ref{fig:answer_sg}(b). Fifteen curves represent three independent runs with different random seeds for each of the five models.}
    \label{fig:evidence_sg}
\end{figure}

Figure~\ref{fig:evidence_sg} shows the model evidence for each of the five models as a function of the hyperparameter $\beta$. We plot the results of three independent runs with different random seeds for each of the five models (15 curves in total). The three curves for each model are indistinguishable on the scale of the figure, indicating the high statistical accuracy. The inset in Fig.~\ref{fig:evidence_sg} shows a magnified view of the region $0.1\le\beta\le 1$ and $-650\le \log P(Y;\beta) \le -540$, indicated by the dashed rectangle. As expected, the super-Gaussian model has the largest peak value of the model evidence and is therefore objectively identified as the most plausible of the five candidates. The evidence peak at $\beta\sim0.5$ further supports the selection of the super-Gaussian model. Among the multi-Maxwellian models, the triple model has the largest model evidence, which also peaks at $\beta\sim 0.5$. Considering also that $\min F(X;Y)/N\sim1$, the triple model appears to provide a reasonable approximation to the ground-truth distribution function. The similar $\min F(X;Y)/N$ values for the triple, quadruple, and super-Gaussian models indicate comparable best-fit quality. The parameter space integration, $\int dX$, in Eq.~\eqref{eq:model_evidence} penalizes additional model complexity, and the model evidence therefore favors the model with the fewest parameters that adequately explains the data.
As shown in Figs.~\ref{fig:comp_fit_sg}(f)--\ref{fig:comp_fit_sg}(i), the distribution functions estimated from the triple and quadruple models closely resemble the ground truth. Because the triple model uses fewer parameters while providing a reconstruction similar to that of the quadruple model, the model evidence favors the triple model over the quadruple model, thereby avoiding overfitting due to unnecessary model complexity.
These results demonstrate that combining a flexible multi-Maxwellian representation with model selection based on the model evidence provides a practical approach to analyzing distribution functions whose exact functional forms are not known in advance. The model selection framework favors a compact model when an appropriate functional form is available and can otherwise provide a parsimonious multi-Maxwellian approximation supported by the data.

\section{Discussion and summary}\label{sec:summary}

\begin{figure}
    \includegraphics[width=\hsize]{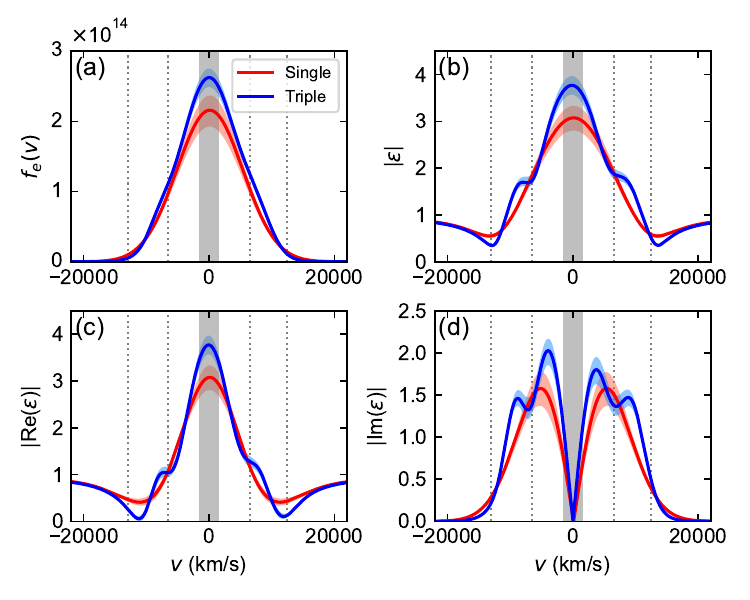}
    \caption{(a) Estimated electron distribution function of single and triple models in the analysis of the spectrum shown in Fig.~\ref{fig:answer}(b). (b) Absolute value, (c) real part, (d) imaginary part of dielectric response function.}
    \label{fig:susceptibility}
\end{figure}

The large ensemble of replicas generated in Bayesian inference enables the calculation of condifence intervals not only for the estimated spectra but also for the underlying distribution functions. This allows us the quantitative comparison of distribution functions across different models. 
Figure~\ref{fig:susceptibility}(a) shows the estimated electron distribution functions of single and triple models, along with their 95\% condifence interval at the optimal $\beta$. The gray background and vertical dotted lines are the same as those in Fig.~\ref{fig:answer}(a). It is clear that no local maxima or ``bump'' structures are observed in the distribution function at the phase velocities corresponding to the additional spectral peaks. Consequently, the additional spectral peaks must be attributed to modifications of the dielectric response function, according to Eq.~\eqref{eq:form} \cite{sakai20pop,sakai23pop}. 
We plot $|\varepsilon|$, $|\mathrm{Re}(\varepsilon)|$, and $|\mathrm{Im}(\varepsilon)|$ in Figs.~\ref{fig:susceptibility}(b)--\ref{fig:susceptibility}(d), respectively. As shown in Fig.~\ref{fig:susceptibility}(b), the dielectric response function of the triple model exhibits distinct dips that give rise to the additional spectral peaks. The real and imaginary parts of $\varepsilon$ characterize the dispersion relation and the Landau damping rate of density fluctuations, respectively. Both the real and imaginary parts in Figs.~\ref{fig:susceptibility}(c) and \ref{fig:susceptibility}(d) exhibit decrease at the phase velocities of additional spectral peaks, reducing $|\varepsilon|$. Since the electron feature is approximated as $S(k,\omega)\sim(2\pi/k)|1/\varepsilon|^2f_e(\omega/k)$, the reduction in $|\varepsilon|$ enhances the spectral response. The changes in the real and imaginary parts indicate modifications to the dispersion relation and reduced Landau damping, respectively. The linear dispersion analysis in Appendix~\ref{sec:dispersion} directly evaluates the wave modes and damping rates. The contribution of the modified dispersion relation contrasts with the discussion by \citet{zhang23nphys}, which attributed the formation of such peaks solely to a reduced Landau damping rate. It is therefore essential to characterize the dielectric response function for a robust physical interpretation of observed data. In this regard, uncertainty quantification via Bayesian inference significantly facilitates the physical interpretation of CTS spectra.

In this study, we use the Penrose criterion to exclude distribution functions that are linearly unstable to electrostatic instabilities \cite{penrose60pof}. This conservative restriction ensures consistency with the quasi-steady state assumption underlying the dynamic structure factor. Because the ground-truth distribution function is linearly stable and unstable replicas are excluded, we neglect temporal evolution during the CTS integration time. For experimental data analysis, however, strict linear stability is not always necessary. Let $\gamma$ denote the linear growth rate and $\tau$ the diagnostic integration time. When $\gamma\tau \ll1$, the distribution function changes negligibly during the measurement, and the quasi-steady state approximation remains valid. When $\gamma\tau\sim1$, temporal evolution can modify the time-integrated spectrum, requiring a time-dependent calculation of the scattered spectrum \cite{sakai20pop,sakai23pop,zhang23nphys,sato26pop}. When $\gamma\tau\gg1$, the instability rapidly depletes the free energy that drives the growth and saturates well within the integration time. After saturation, the plasma can remain in a quasi-steady state for most of the measurement, allowing the time-integrated spectrum to be described using the dynamic structure factor \cite{sakai20pop}. 
For applications to experimental data, the present hard stability constraint could be replaced by a condition based on the growth rate and diagnostic integration time. For a sampled distribution function for which the Penrose criterion predicts an instability, we could solve the linear dispersion relation for complex frequency near the phase velocity associated with the relevant local minimum \cite{kato10pop}. A replica could be retained when $\gamma\tau\ll1$ because temporal evolution is negligible during the measurement. A replica with $\gamma\tau\gtrsim1$ should be excluded from a forward model based on the stationary dynamic structure factor because the sampled distribution function cannot remain stationary over the integration time. Without this constraint, the inverse problem analysis sometimes favors solutions that reproduce the measured spectrum but correspond to unstable distribution functions that are not physically relevant.

As discussed in Secs.~\ref{sec:model_evidence} and \ref{sec:superGaussian}, we demonstrated a model selection framework based on model evidence that eliminates arbitrariness. This approach is particularly effective for resolving rarefied velocity-space structures with low signal-to-noise ratios, such as reflected ions in collisionless shocks \cite{yamazaki22pre,matsukiyo22pre} and outflows during magnetic reconnections \cite{sakai22srep,morita22pre,ji24pop,zhang23nphys}. 
However, there is a wide variety of models describing non-Maxwellian distribution functions. Since each model is currently analyzed independently, comparing model evidence requires repeating the inference process for every candidate model using the same spectral dataset. Given that standard CTS diagnostics involve spatial or temporal profiles, applying this model selection process to all coordinates requires substantial computational resources.
A promising solution for accelerating the analysis is leveraging the replica resampling feature in the PAMC method, which resample replicas based on their respective Neal-Jarzynski weights during the annealing process \cite{hukushima03aipp,motoyama22cpc}. By initializing replicas of different models within a single PAMC run and performing resampling according to their respective probability, the population of the most probable model is expected to become dominant at $\beta$ with the largest model evidence. The implementation and validation of this approach are left for future work.

While this study focuses on the inverse problem analysis of the electron feature in CTS spectra, this can be easily extended to the ion feature. In the ion feature analysis, both non-Maxwellian distribution functions \cite{morita22pre} and multiple ion species effects \cite{glenzer96prl} modify the spectral response; therefore, distinguishing between these two underlying physics is critical. We anticipate that the model selection framework presented here will effectively differentiate between non-Maxwellian distribution functions and multiple ion species effects. Furthermore, simultaneous measurements of both electron and ion features could provide additional experimental constraints to refine this identification \cite{nakano26pfr}. 

In summary, we have developed a framework for the Bayesian inference of non-Maxwellian distribution functions from collective Thomson scattering (CTS) spectra. 
Using synthetic spectra generated from known ground-truth distribution functions, we compare multiple candidate models with different numbers of parameters and identify the most plausible model based on model evidence. When the candidate models include the ground-truth model, the model evidence favors the ground-truth model. Without the ground-truth model as a candidate, a sufficiently flexible candidate model approximates the overall shape of the ground-truth distribution function, while the model evidence favors the candidate model with the fewest parameters required to adequately explain the data.
Through the posterior probability density function, we obtain the characteristics of the estimated distribution function. The observed spectral features are physically interpreted as a combination of changes in the dispersion relation and reduced Landau damping. This technique removes both the arbitrariness of model selection and the dependence on initial guesses in data analysis, enabling the identification of distribution functions solely from the observed CTS spectra.

\begin{acknowledgments}
The computation in this work has been done using the facilities of the Supercomputer Center, the Institute for Solid State Physics, the University of Tokyo (2023-Ca-0122, 2024-Ca-0120, and 2025-Ca-0130) and the Fugaku supercomputer through the HPCI System Research Projects (Project ID: hp220248, hp230304, hp240304, hp250295).
The authors thank H.~Ji, L.~Gao, K.~Yoshimi, and K.~Hukushima for helpful comments. 
The authors are supported by JST Moonshot R\&D Program (Grant No. JPMJMS24A3) for the data analysis using ODAT-SE. 
K.S. is supported by JSPS KAKENHI (Grant Nos. JP21J20499 and JP24K17029), 
by the NINS program of Promoting Research by Networking among Institutions (Grant Nos. 01422301 and 01412302),
and by Foundation of Public Interest of Tatematsu.
\end{acknowledgments}

\appendix

\section{Linear dispersion analysis}\label{sec:dispersion}

\begin{figure}
    \includegraphics[width=\hsize]{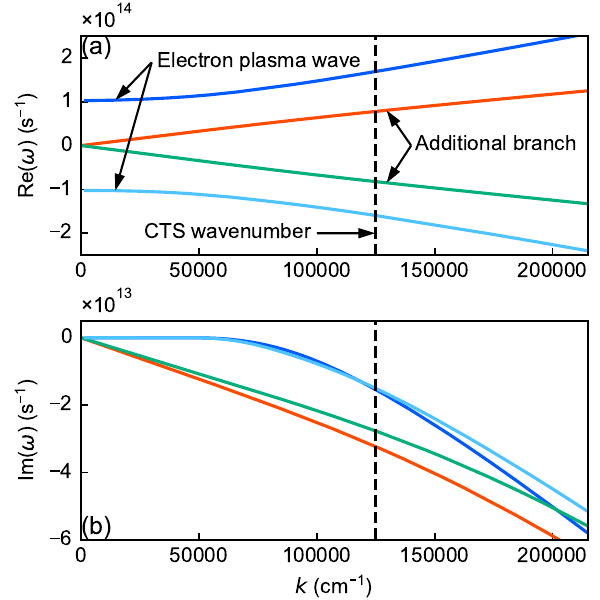}
    \caption{(a) Real and (b) imaginary parts of linear dispersion relation using the electron distribution function in Fig.~\ref{fig:answer}(a).}
    \label{fig:dispersion}
\end{figure}
We performed the inverse problem analysis using the spectrum assuming the electron distribution function in Fig.~\ref{fig:answer}(a). Since the distribution function is single-humped, it is linearly stable to electrostatic instabilities \cite{penrose60pof}. We calculated the linear dispersion relation to quantitatively evaluate the wave modes and the stability \cite{kato10pop}. The ion susceptibility is neglected in this analysis. Figure~\ref{fig:dispersion}(a) shows two additional branches besides the two electron plasma wave branches. At the wavenumber of CTS measurement (vertical dashed line), the phase velocities of the additional branches are consistent with the velocity associated with the additional peaks in the CTS spectrum in Fig.~\ref{fig:answer}(b). The imaginary parts of all branches in Fig.~\ref{fig:dispersion}(b) are negative, indicating that no linearly unstable modes are present. At CTS wavenumber, the additional and electron plasma wave branches have comparable damping rates, with a ratio of approximately two. The comparable damping allows the additional modes to produce observable peaks in the CTS spectrum.

\bibliography{ref.bib}

@book{froula11,
    title={Plasma Scattering of Electromagnetic Radiation: Theory and Measurement Techniques},
    author={Froula, Dustin H. and Glenzer, Siegfried H. and Luhmann, Jr., Neville C. and Sheffield, John},
    edition = {2nd},
    year={2011},
    publisher={Academic Press},
    address = {Amsterdam},
    doi={10.1016/C2009-0-20048-1},
}

@article{henchen18prl,
  title = {Observation of Nonlocal Heat Flux Using {Thomson} Scattering},
  author = {Henchen, R. J. and Sherlock, M. and Rozmus, W. and Katz, J. and Cao, D. and Palastro, J. P. and Froula, D. H.},
  journal = {Physical Review Letters},
  volume = {121},
  issue = {12},
  pages = {125001},
  numpages = {6},
  year = {2018},
  month = {Sep},
  publisher = {American Physical Society},
  doi = {10.1103/PhysRevLett.121.125001},
}

@article{sakai20pop,
    author = {Sakai,K.  and Isayama,S.  and Bolouki,N.  and Habibi,M. S.  and Liu,Y. L.  and Hsieh,Y. H.  and Chu,H. H.  and Wang,J.  and Chen,S. H.  and Morita,T.  and Tomita,K.  and Yamazaki,R.  and Sakawa,Y.  and Matsukiyo,S.  and Kuramitsu,Y. },
    title = {Collective {Thomson} scattering in non-equilibrium laser produced two-stream plasmas},
    journal = {Physics of Plasmas},
    volume = {27},
    number = {10},
    pages = {103104},
    year = {2020},
    doi = {10.1063/5.0011935},
}

@article{fiuza20nphys,
  author          = {Fiuza, F. and Swadling, G. F. and Grassi, A. and Rinderknecht, H. G. and Higginson, D. P. and Ryutov, D. D. and Bruulsema, C. and Drake, R. P. and Funk, S. and Glenzer, S. and Gregori, G. and Li, C. K. and Pollock, B. B. and Remington, B. A. and Ross, J. S. and Rozmus, W. and Sakawa, Y. and Spitkovsky, A. and Wilks, S. and Park, H.-S.},
  title           = {Electron acceleration in laboratory-produced turbulent collisionless shocks},
  journal         = {Nature Physics},
  volume          = {16},
  pages           = {916–920},
  year            = {2020},
  doi             = {10.1038/s41567-020-0919-4},
}

@article{yamazaki22pre,
  title = {High-power laser experiment forming a supercritical collisionless shock in a magnetized uniform plasma at rest},
  author = {Yamazaki, R. and Matsukiyo, S. and Morita, T. and Tanaka, S. J. and Umeda, T. and Aihara, K. and Edamoto, M. and Egashira, S. and Hatsuyama, R. and Higuchi, T. and Hihara, T. and Horie, Y. and Hoshino, M. and Ishii, A. and Ishizaka, N. and Itadani, Y. and Izumi, T. and Kambayashi, S. and Kakuchi, S. and Katsuki, N. and Kawamura, R. and Kawamura, Y. and Kisaka, S. and Kojima, T. and Konuma, A. and Kumar, R. and Minami, T. and Miyata, I. and Moritaka, T. and Murakami, Y. and Nagashima, K. and Nakagawa, Y. and Nishimoto, T. and Nishioka, Y. and Ohira, Y. and Ohnishi, N. and Ota, M. and Ozaki, N. and Sano, T. and Sakai, K. and Sei, S. and Shiota, J. and Shoji, Y. and Sugiyama, K. and Suzuki, D. and Takagi, M. and Toda, H. and Tomita, S. and Tomiya, S. and Yoneda, H. and Takezaki, T. and Tomita, K. and Kuramitsu, Y. and Sakawa, Y.},
  journal = {Physical Review E},
  volume = {105},
  issue = {2},
  pages = {025203},
  numpages = {16},
  year = {2022},
  month = {Feb},
  publisher = {American Physical Society},
  doi = {10.1103/PhysRevE.105.025203},
}

@article{schaeffer19prl,
  title = {Direct Observations of Particle Dynamics in Magnetized Collisionless Shock Precursors in Laser-Produced Plasmas},
  author = {Schaeffer, D. B. and Fox, W. and Follett, R. K. and Fiksel, G. and Li, C. K. and Matteucci, J. and Bhattacharjee, A. and Germaschewski, K.},
  journal = {Physical Review Letters},
  volume = {122},
  issue = {24},
  pages = {245001},
  numpages = {6},
  year = {2019},
  month = {Jun},
  publisher = {American Physical Society},
  doi = {10.1103/PhysRevLett.122.245001},
}

@article{sakai22srep,
  author          = {Sakai, K. and Moritaka, T. and Morita, T. and Tomita, K. and Minami, T. and Nishimoto, T. and Egashira, S. and Ota, M. and Sakawa, Y. and Ozaki, N. and Kodama, R. and Kojima, T. and Takezaki, T. and Yamazaki, R. and Tanaka, S. J. and Aihara, K. and Koenig, M. and Albertazzi, B. and Mabey, P. and Woolsey, N. and Matsukiyo, S. and Takabe, H. and Hoshino, M. and Kuramitsu, Y.},
  journal         = {Scientific Reports},
  title           = {Direct observations of pure electron outflow in magnetic reconnection},
  year            = {2022},
  month           = {Jun},
  number          = {1},
  volume          = {12},
  pages           = {10921},
  issn            = {2045-2322},
  doi             = {10.1038/s41598-022-14582-3},
}

@article{milder21prl,
  title = {Measurements of Non-{Maxwellian} Electron Distribution Functions and Their Effect on Laser Heating},
  author = {Milder, A. L. and Katz, J. and Boni, R. and Palastro, J. P. and Sherlock, M. and Rozmus, W. and Froula, D. H.},
  journal = {Physical Review Letters},
  volume = {127},
  issue = {1},
  pages = {015001},
  numpages = {7},
  year = {2021},
  month = {Jun},
  publisher = {American Physical Society},
  doi = {10.1103/PhysRevLett.127.015001},
}

@article{kato10pop,
author = {Kato,Tsunehiko N.  and Takabe,Hideaki },
title = {Electrostatic and electromagnetic instabilities associated with electrostatic shocks: Two-dimensional particle-in-cell simulation},
journal = {Physics of Plasmas},
volume = {17},
number = {3},
pages = {032114},
year = {2010},
doi = {10.1063/1.3372138},
}

@article{takabe21hpl, 
  title={Recent progress of laboratory astrophysics with intense lasers}, 
  volume={9}, 
  DOI={10.1017/hpl.2021.35}, 
  journal={High Power Laser Science and Engineering}, 
  publisher={Cambridge University Press}, 
  author={Takabe, Hideaki and Kuramitsu, Yasuhiro}, 
  year={2021}, 
  pages={e49}
}

@article{turnbull20nphys,
  author          = {Turnbull, David and Colaïtis, Arnaud and Hansen, Aaron M. and Milder, Avram L. and Palastro, John P. and Katz, Joseph and Dorrer, Christophe and Kruschwitz, Brian E. and Strozzi, David J. and Froula, Dustin H.},
  journal         = {Nature Physics},
  number          = {181-185},
  title           = {Impact of the {Langdon} effect on crossed-beam energy transfer},
  volume          = {16},
  pages           = {181–185},
  year            = {2020},
  doi             = {10.1038/s41567-019-0725-z},
}

@article{matsukiyo06jgr,
  author = {Matsukiyo, S. and Scholer, M.},
  title = {On microinstabilities in the foot of high {Mach} number perpendicular shocks},
  journal = {Journal of Geophysical Research: Space Physics},
  volume = {111},
  number = {A6},
  pages = {A06104},
  doi = {10.1029/2005JA011409},
  year = {2006},
}

@article{morita22pre,
  title = {Detection of current-sheet and bipolar ion flows in a self-generated antiparallel magnetic field of laser-produced plasmas for magnetic reconnection research},
  author = {Morita, T. and Kojima, T. and Matsuo, S. and Matsukiyo, S. and Isayama, S. and Yamazaki, R. and Tanaka, S. J. and Aihara, K. and Sato, Y. and Shiota, J. and Pan, Y. and Tomita, K. and Takezaki, T. and Kuramitsu, Y. and Sakai, K. and Egashira, S. and Ishihara, H. and Kuramoto, O. and Matsumoto, Y. and Maeda, K. and Sakawa, Y.},
  journal = {Physical Review E},
  volume = {106},
  issue = {5},
  pages = {055207},
  numpages = {12},
  year = {2022},
  month = {Nov},
  publisher = {American Physical Society},
  doi = {10.1103/PhysRevE.106.055207},
}

@article{yao21nphys,
  author          = {Yao, W. and Fazzini, A. and Chen, S. N. and Burdonov, K. and Antici, P. and Béard, J. and Bolaños, S. and Ciardi, A. and Diab, R. and Filippov, E. D. and Kisyov, S. and Lelasseux, V. and Miceli, M. and Moreno, Q. and Nastasa, V. and Orlando, S. and Pikuz, S. and Popescu, D. C. and Revet, G. and Ribeyre, X. and d'Humières, E. and Fuchs, J.},
  journal         = {Nature Physics},
  volume          = {17},
  pages           = {1177–1182},
  title           = {Laboratory evidence for proton energization by collisionless shock surfing},
  year            = {2021},
  doi             = {10.1038/s41567-021-01325-w},
}

@article{fujimoto14grl,
author = {Fujimoto, Keizo},
title = {Wave activities in separatrix regions of magnetic reconnection},
journal = {Geophysical Research Letters},
volume = {41},
number = {8},
pages = {2721–2728},
doi = {10.1002/2014GL059893},
year = {2014}
}

@article{sakai23pop,
author = {Sakai,K.  and Nishimoto,T.  and Isayama,S.  and Matsukiyo,S.  and Kuramitsu,Y. },
title = {Ion-acoustic feature of collective {Thomson} scattering in non-equilibrium two-stream plasmas},
journal = {Physics of Plasmas},
volume = {30},
number = {1},
pages = {012105},
year = {2023},
doi = {10.1063/5.0117812},
}

@article{matsukiyo22pre,
  title = {High-power laser experiment on developing supercritical shock propagating in homogeneously magnetized plasma of ambient gas origin},
  author = {Matsukiyo, S. and Yamazaki, R. and Morita, T. and Tomita, K. and Kuramitsu, Y. and Sano, T. and Tanaka, S. J. and Takezaki, T. and Isayama, S. and Higuchi, T. and Murakami, H. and Horie, Y. and Katsuki, N. and Hatsuyama, R. and Edamoto, M. and Nishioka, H. and Takagi, M. and Kojima, T. and Tomita, S. and Ishizaka, N. and Kakuchi, S. and Sei, S. and Sugiyama, K. and Aihara, K. and Kambayashi, S. and Ota, M. and Egashira, S. and Izumi, T. and Minami, T. and Nakagawa, Y. and Sakai, K. and Iwamoto, M. and Ozaki, N. and Sakawa, Y.},
  journal = {Physical Review E},
  volume = {106},
  issue = {2},
  pages = {025205},
  numpages = {7},
  year = {2022},
  month = {Aug},
  publisher = {American Physical Society},
  doi = {10.1103/PhysRevE.106.025205},
}

@Article{zhang23nphys,
author={Zhang, Shu and Chien, Abraham and Gao, Lan and Ji, Hantao and Blackman, Eric G. and Follett, Russ and Froula, Dustin H. and Katz, Joseph and Li, Chikang and Birkel, Andrew and Petrasso, Richard and Moody, John and Chen, Hui},
title={Ion and electron acoustic bursts during anti-parallel magnetic reconnection driven by lasers},
journal={Nature Physics},
year={2023},
month={Jun},
day={01},
volume={19},
number={6},
pages={909–916},
issn={1745-2481},
doi={10.1038/s41567-023-01972-1},
}

@article{motoyama22cpc,
title = {Data-analysis software framework {2DMAT} and its application to experimental measurements for two-dimensional material structures},
journal = {Computer Physics Communications},
volume = {280},
pages = {108465},
year = {2022},
issn = {0010-4655},
doi = {10.1016/j.cpc.2022.108465},
author = {Yuichi Motoyama and Kazuyoshi Yoshimi and Izumi Mochizuki and Harumichi Iwamoto and Hayato Ichinose and Takeo Hoshi},
}

@article{ji24pop,
    author = {Ji, H. and Gao, L. and Pomraning, G. and Sakai, K. and Guo, F. and Li, X. and Stanier, A. and Milder, A. and Follett, R. K. and Fiksel, G. and Blackman, E. G. and Chien, A. and Zhang, S.},
    title = {Study of magnetic reconnection at low-$\beta$ using laser-powered capacitor coils},
    journal = {Physics of Plasmas},
    volume = {31},
    number = {10},
    pages = {102112},
    year = {2024},
    month = {10},
    issn = {1070-664X},
    doi = {10.1063/5.0223922},
}

@Article{escalona23srep,
author={Escalona, M. and Valenzuela, J. C. and Avaria, G. and Veloso, F. and Wyndham, E. S.},
title={Bayesian inference of plasma parameters from collective {Thomson} scattering technique on a gas-puff near stagnation},
journal={Scientific Reports},
year={2023},
month={Aug},
day={10},
volume={13},
number={1},
pages={13002},
issn={2045-2322},
doi={10.1038/s41598-023-40014-x},
}

@article{foo23aipa,
    author = {Foo, B. C. and Schaeffer, D. B. and Heuer, P. V.},
    title = {Recovering non-{Maxwellian} particle velocity distribution functions from collective {Thomson}-scattered spectra},
    journal = {AIP Advances},
    volume = {13},
    number = {11},
    pages = {115328},
    year = {2023},
    month = {11},
    issn = {2158-3226},
    doi = {10.1063/5.0169393},
}

@article{friel12sn,
author = {Friel, Nial and Wyse, Jason},
title = {Estimating the evidence – a review},
journal = {Statistica Neerlandica},
volume = {66},
number = {3},
pages = {288–308},
doi = {10.1111/j.1467-9574.2011.00515.x},
year = {2012}
}

@article{hukushima03aipp,
    author = {Hukushima, K. and Iba, Y.},
    title = {Population Annealing and Its Application to a Spin Glass},
    journal = {AIP Conference Proceedings},
    volume = {690},
    number = {1},
    pages = {200–206},
    year = {2003},
    month = {11},
    issn = {0094-243X},
    doi = {10.1063/1.1632130},
}

@article{scholer03jgr,
author = {Scholer, Manfred and Shinohara, Iku and Matsukiyo, Shuichi},
title = {Quasi-perpendicular shocks: Length scale of the cross-shock potential, shock reformation, and implication for shock surfing},
journal = {Journal of Geophysical Research: Space Physics},
volume = {108},
number = {A1},
pages = {SSH 4-1-SSH 4-11},
doi = {10.1029/2002JA009515},
year = {2003}
}

@article{sato26pop,
    author = {Sato, Yuma and Matsukiyo, Shuichi},
    title = {Two-dimensional {PIC} simulation of collective {Thomson} scattering in a beam-plasma system},
    journal = {Physics of Plasmas},
    volume = {33},
    number = {1},
    pages = {013901},
    year = {2026},
    month = {01},
    issn = {1070-664X},
    doi = {10.1063/5.0304809},
}

@article{glenzer96prl,
  title = {Observation of Two Ion-Acoustic Waves in a Two-Species Laser-Produced Plasma with {Thomson} Scattering},
  author = {Glenzer, S. H. and Back, C. A. and Estabrook, K. G. and Wallace, R. and Baker, K. and MacGowan, B. J. and Hammel, B. A. and Cid, R. E. and De Groot, J. S.},
  journal = {Physical Review Letters},
  volume = {77},
  issue = {8},
  pages = {1496–1499},
  numpages = {0},
  year = {1996},
  month = {Aug},
  publisher = {American Physical Society},
  doi = {10.1103/PhysRevLett.77.1496},
}

@article{chapman11prl,
  title = {Analysis of {Thomson} Scattering from Nonequilibrium Plasmas},
  author = {Chapman, D. A. and Gericke, D. O.},
  journal = {Physical Review Letters},
  volume = {107},
  issue = {16},
  pages = {165004},
  numpages = {5},
  year = {2011},
  month = {Oct},
  publisher = {American Physical Society},
  doi = {10.1103/PhysRevLett.107.165004},
}

@article{sakai26ppcf,
doi = {10.1088/1361-6587/ae37a4},
year = {2026},
month = {jan},
publisher = {IOP Publishing},
volume = {68},
number = {1},
pages = {015034},
author = {Sakai, Kentaro and Tomita, Kentaro and Hoshi, Takeo and Nakano, Akito and Goto, Motoshi and Nagaoka, Kenichi and Yasuhara, Ryo},
title = {Conceptual design of {Thomson} scattering system with high wavelength resolution in magnetically confined plasmas for electron phase-space measurements},
journal = {Plasma Physics and Controlled Fusion},
}

@article{penrose60pof,
    author = {Penrose, Oliver},
    title = {Electrostatic Instabilities of a Uniform Non-{Maxwellian} Plasma},
    journal = {The Physics of Fluids},
    volume = {3},
    number = {2},
    pages = {258–265},
    year = {1960},
    month = {03},
    issn = {0031-9171},
    doi = {10.1063/1.1706024},
}

@article{pavan11pop,
    author = {Pavan, J. and Yoon, P. H. and Umeda, T.},
    title = {Quasilinear theory and simulation of {Buneman} instability},
    journal = {Physics of Plasmas},
    volume = {18},
    number = {4},
    pages = {042307},
    year = {2011},
    month = {04},
    issn = {1070-664X},
    doi = {10.1063/1.3574359},
}

@article{nakano26pfr,
  author       = {Nakano, Akito and Sakai, Kentaro and Ji, Hantao and Gao, Lan and Russell, Brandon K. and Hoshi, Takeo},
  title        = {Two-component {Bayesian} modeling in the data analysis of collective {Thomson} scattering},
  journal = {Plasma and Fusion Research, in press},
  Year = {2026},
  Eprint = {arXiv:2610.32033},
}

\end{document}